\documentclass[conference]{IEEEtran}
\usepackage{graphicx}

\usepackage{cite}

\ifCLASSINFOpdf
\else
\fi
\usepackage{graphicx}

\usepackage[cmex10]{amsmath}
\usepackage{mathtools}
\usepackage{amsfonts}
\usepackage{amssymb}
\usepackage{subfig}

\usepackage{epstopdf}

\begin{document}
%
\title{A blockchain based Secure and Trusted framework for Information Propagation on Online Social Networks  \\
}

\author{\IEEEauthorblockN{Md Arquam}
\IEEEauthorblockA{\textit{Computer Science and Engineering } \\
\textit{National Institute of Technology Delhi}\\
New Delhi, India \\
arquam@nitdelhi.ac.in}
\and
\IEEEauthorblockN{Anurag Singh}
\IEEEauthorblockA{\textit{Computer Science and Engineering} \\
\textit{National Institute of Technology Delhi}\\
New Delhi, India \\
anuragsg@nitdelhi.ac.in}
\and
\IEEEauthorblockN{Rajesh Sharma}
\IEEEauthorblockA{\textit{Institute of Computer Science} \\
\textit{University of Tartu}\\
Tartu, Estonia \\
rajesh.sharma@ut.ee}

}

\maketitle


\maketitle
\begin{abstract}
The online social networks facilitate naturally for the users to share information. On these platforms, each user shares information based on his or her interests. The particular information being shared by a user may be legitimate or fake. Sometimes a misinformation, propagated by users and group can create chaos or in some cases, might leads to cases of riots. Nowadays the third party like ALT news and Cobrapost check the
information authenticity, but it takes too much time to validate the news. Therefore, a robust and new system is required to check the information authenticity within the network, to stop the propagation of misinformation.

In this paper, we propose a blockchain based framework for sharing the information securely at the peer level. In the blockchain model, a chain is created by combining blocks of information. Each node of network propagates the information based on its credibility to its peer nodes. The credibility of a node will
vary according to the respective information. Trust is calculated between sender and receiver in two ways:(i) Local trust used for sharing information at the peer level and (ii) global trust is used for a credibility check of each user in the network. We evaluate our framework using real dataset derived from Facebook. Our approach achieves an accuracy of 83\% which shows the effectiveness of our proposed framework.

\end{abstract}

\begin{IEEEkeywords}
Blockchain , Information Dynamics, Social Network, 
\end{IEEEkeywords}

\section{Introduction}

Information spreading among humans is natural phenomena. A decade back it happened mainly through offline interactions. However, with the advancement of the web, information spreads mainly through online social media and that too very fast as people are densely connected with each other~\cite{matsubara2012rise}. An information spreading on a network may be legitimate or might not be correct. An incorrect information may be termed as rumor related to public interest~\cite{peterson1951rumor}. Nowadays social network, like Facebook, Twitter is very common for communication among people to do collaborative action, for example, the Arab spring uprising and London riots. Researchers already discussed the quick spreading of information propagation due to the structure of the social network~\cite{doerr2012rumors}.

Researchers studied the dynamics of rumor spreading considering both the modelling technique as well as the mechanism to avoid the rumor spreading~\cite{jin2013epidemiological}. Therefore, the relationship between the dynamics of information and the structure of the underlying network is crucial in many real cases, e.g., the spreading of worms in a computer network (eg., ransom virus on technological networks), viruses in a human population (eg., zika virus spreading in human) , information propagation in the online social network. Since rumor is generated and transmitted over extended periods of time, therefore, it is important to find out the authenticity of information. 

The blockchain technology is developed for the financial transaction of bitcoin with trusted and secured contract between two communicating parties at the peer level.  This method motivates us to modify the blockchain technology for the information dynamics by considering trusted-contract based propagation on the social network.  In addition, blockchain based contract helps to find the source of information generation. Hence, it is interesting to develop a method to check the authenticity of information through blockchain technology in the social network to prevent the rumour spreading.

\subsection{Motivation}

Earlier research put in force the focus on both, that is the development of information propagation model as well as rumor spreading, however, treating these two topics separately \cite{nekovee2007theory, moreno2004dynamics}. Therefore, intensive research is required for information dynamics in which, correct information should be propagated as well as misinformation or false information should be blocked to stop the chaos. The verity of the social network in collaboration with the emanation of information transmission media, researchers are not able to propose an effective method to abolish the rumor dissemination implicitly in the network. 

Most of the research on information dynamics is done by considering either rumors propagation~\cite{nekovee2007theory} or only positive information propagation~\cite{walter2008model, acemoglu2011opinion} separately. There are no criteria being set to find out, how to block the misinformation within the network by using network properties. Also, the work is missing with respect to exploration of the impact of correct as well as incorrect information on trust simultaneously.

Therefore, a new technique or method is required to solve the problem of information dynamics considering verification and authentication of information, near to the initial period of the starting of the information, in the social network so that immediate action should be taken to remove the unverified information. In addition, it is also important to find the source of information blast to take proper action against the misinformation originators.

\subsection{Research Questions}
There are many research question with the existing social networks as to find the trusted friends, connections(links), authenticity check of shared information and collected information. Anyone can share information with its friends. Most of the social networking sites, for example, Facebook and Twitter rely on traditional methods to find a trusted mutual link based on a mutual friend relationship. This method is not appropriate to find trustworthy friends. The various research questions in social networks are listed as follows:
\begin{enumerate}
\item[RQ 1.] \textbf{Trustworthy Large-Scale Social Networks Evaluation:} Trust is an important basis for interactions among nodes in large networks. However, the scale and stochasticity of such networks make it impossible for each party to make trusting with each other. Due to this reason, the users in large networks must share information about the trust. However, they should not trust such shared information automatically.
\item[RQ 2.] \textbf{Data Privacy Preserving:} Every day, more and more social network data has been published out, so preserving of data privacy in the social network becomes crucial. If an adversary has some knowledge about the neighbours of a target victim and the relationship among the neighbours,  an adversary may attack the privacy of some victims easily.
\item[RQ 3.] \textbf{Friend Recommendation :} Existing social network suggest friends to users, based on their social network activity, which may not be the suitable to reflect a user's preferences on friend selection in real life. 
\item[RQ 4.] \textbf{Trust Based user Credibility:} First choice of people is social media to get news and information. Therefore, it becomes challenging for a user to trust on which information is credible or not. They require to find ways to check the credibility of the information. This problem becomes much more critical when the source of the information is not known to the other.
\end{enumerate}
 
\subsection{Contribution}

In this paper, we study the authenticity of the information, propagating on the social network. To check the authenticity of information, we apply the blockchain technology by considering the Facebook social network. The blockchain method propagates the information to it's trusted peer nodes. These peer nodes check the credibility of the message generating nodes before accepting the information. Based on the credibility, information is classified as correct or incorrect.  We apply the network parameter of underlying topology to define the smart contract for blockchain technology rather than using the fixed virtual credit for each node. 
By using blockchain technology we also able to find the source of information generating nodes 

The rest of this paper is organized as follows: Section II discusses the current state of arts regarding information dynamics on social network and application of blockchain other than bitcoin. Section III explains the proposed methodology that includes the application of blockchain in information propagation on the social network by cons. Section IV presents the simulation and result analysis. In this section, we have simulated the model to get the result. Finally, Section V describes conclusions and outlines some of our future directions.

\section{Related Work}
In this section, we first present various works related to social networks, including trust on social networks, the concept which is an integral part of our framework (Section \ref{sec:rw:sn}) and then we describe works related to the blockchain technology (Section \ref{sec:rw:bc}).

\subsection{Social networks}\label{sec:rw:sn}
There are many Online Social Networks and every now and then we see the emergence of new online social platforms. The primary reason of the presence of so many social networks is due to the fact that they play a key role in online personal and commercial interaction and most importantly they play an important role to find the source of information and knowledge. Online social networks such as Facebook, Twitter, LinkedIn, and Google+ have become popular, where people around the world get connected and share information with each other~\cite{williamson2010social}.\\

Each social network contains user profiles, a list of connected users called friends and their topic of interests~\cite{nosko2010all}. According to their interest, users share information on these social platforms. Therefore, social networks provide a base to maintain social relationships, to find other users with similar interests, as well as to find the content that has been shared by other users~\cite{mislove2007measurement}. Every person has a certain mindset and topic of interest, according to their ease, that may or may not be changed. In addition, each user in a social network creates trust with its neighbours at peer level before sharing the information. Trust is a measure of confidence in social networks and it provides the information about the neighbours with whom, what type of information one share/accept with others~\cite{walter2008model}. Therefore, trust in social networks has attracted a lot of attention for opinion~\cite{acemoglu2011opinion}, recommendation~\cite{walter2008model}. Hence, the trust may be used in any system effectively, e.g. computer science, cognitive sciences, sociology, and economics. It is important to calculate the trust of a node when it will contact one or more neighbours. 

Each node is connected to its peer nodes based on some common interests in the social network. The information is propagated from one node to its peer nodes and further next level of neighbours. In this way, information is propagated in the social network based on connectivity~\cite{nekovee2007theory}. Hence, the study of underlying network topology is also important. Therefore, the study of social network includes both the dynamics on the network as well as underlying network topology~\cite{moreno2004dynamics}. Most of the social networks are the scale-free network~\cite{li2005towards}.

 

\subsection{The Blockchain Technology}\label{sec:rw:bc}
A blockchain is a collection of blocks connected in sequential order with respect to time. Each block carries the respective record and timestamp. This chain is created for the propagation of records of a node with other connected nodes in a peer-to-peer network. The design of blockchain prevents the modification of data and validates the records~\cite{nakamoto2008bitcoin}.

Researchers have investigated the blockchain protocol to create a decentralized network. In such a type of network, there is no need of the third party for authentication and validation. An automated access control manager is used instead of a third party. That access manager is enabled by distributed blockchain system ~\cite{zyskind2015decentralizing}. Also, researchers have applied the concept of blockchain in supply chain management to improve the quality. It has also shown that blockchain technology provides both utilities and consumers benefit by recording and validating the information ~\cite{chen2017blockchain}. The utilization of cloud storage can be increased by a combination of blockchain and IOT (Internet of Things)~\cite{shafagh2017towards}.  The other area of applications of the blockchain other than online transaction is identity management and notarization~\cite{zyskind2015decentralizing}.



\section{Proposed Methodology}
In this section, first, we describe the network model as a decentralized network and evaluate the trust value by considering network parameter like the degree of a node of a given network. Trust in the network may be defined as the agreement to believe that some node (user in a social network) is good and honest and will not harm you, or that something is safe and reliable. Based on this trust, we define the credibility score of each node according to the message type. Credibility may be called as the fact that someone can be believed or trusted. Finally, we explain the blockchain based information propagation.
\subsection{Network Model} \label{network}
In this work, we consider social networks as a decentralized network. 
Let a node \textbf{A} and has limited interests in various topics. For example, \textbf{A} is very good in science and technology related subject, hence, he shares most of the time science-related information to other nodes. Simultaneously,  \textbf{A} is a movie-loving guy also. Then he has a certain connection with other movie-loving users. 
\textbf{A} also talks about religion. Therefore, he has a connection with religious users. it clearly shows the different forms and context of \textbf{A} and \textbf{A} effortlessly glides from one to the other.

We model our network as an undirected graph, $G = (V, E, C, I)$ , where $V$ is set of $n$ vertices, such that $V = {\{V_1,V_2,...V_n\}}$ , each  vertex depicting a node (or user) in the network has a set of limited 
connections with other vertices, which represents its neighbors. $E$ is the set of edges, where each edge $e_{ij}$ $\in$ $E$, connects a vertex $v_i$ with the vertex $v_j$. $C$ represents a credibility set and total 
interests of the network is represented by $I$. Each vertex $v_i$ has a particular credibility $c_i$ $\in$ $C$. It has global impact on network and different from trust and has limited set of interests $I_i$.

\subsection{Trust Relationship}

In the proposed " A blockchain based Secure and Trusted framework for Information Propagation on Online Social Networks" model, we consider trust relationships among the users. The aim of introducing trust among the nodes is to find the suitable user to validate or invalidate the information in order to propagate the information. Each user, $v_i$ keeps track of a trust value $T_{v_i ,v_j}$ with each of its neighboring users, $v_j$. Trust may be considered in two ways, one is private trust (or local trust) between two communicating nodes and, the other is the public trust (or global trust), in which a source node broadcasts the information in the network about the type of information which he has. 


\subsubsection{Local Trust}
It is important to note that local trust exists only among neighbors of a node in the network. Local trust is based on the physical properties of the network, where a node makes connection with similar type of nodes. To make local trust by exploring the structural properties of underlying network topology, pearson correlation coefficient method is used for structural similarity of the nodes. The Pearson coefficient is given by,

\begin{equation}
r_{s,d} = \frac{\sum_k (A_{ik}-\langle A_i \rangle)(A_{jk}-\langle 
A_j \rangle)}{\sqrt{\sum_k (A_{ik}-\langle A_i \rangle)^2}\sqrt{\sum_k 
(A_{jk}-\langle A_j \rangle)^2}}
\end{equation}
where, $r_{s,d}$ is Pearson coefficient,\\
$A_{ik}$ is $k$ number of neighbors of $i$ \\
$A_{jk}$ is $k$ number of neighbors of $j$ \\
$\langle A_{i}\rangle$ is the mean of $i^{th}$ row of the adjacency 
matrix of $A_{i,k}$\\
$\langle A_{j}\rangle$ is the mean of $j^{th}$ row of the adjacency 
matrix of $A_{j,k}$ \\
The value of quantity $r_{s,d}$ lies strictly in the range $ -1 
\leq r_{s,d} \leq 1$.

Based on local trust, the following steps are taken for the information propagation,

\begin{enumerate}
\item Consider a group of seed nodes $V_s$ $\in$ $ V $ wants to propagate information in the network with $V_r$ $\in$ $V$ neighbors. 
\item At each time stamp $t$ when node $V_s$ makes contact to its neighbor nodes $V_r$ to send information. \item Validation of an information i.e. authenticity of information is based on the credibility of the sender node.
\item Once information is validated, it is propagated.
\item When neighbors invalidate the information of the sender, it cannot be propagated further.
\item Once information is validated its credibility is changed in global trust.
\end{enumerate}

\subsubsection{Global Trust}
The global trust is updated with information propagation to record the nodes involved in propagation, the credibility of the nodes and, the type of information being propagated. In other words, it works as the public ledger for all information propagation.  An information propagation with their credibility is used as proof of a contract to all available nodes of the information exchange, which is used for decision-making in local trust.

%
%

Following steps are taken for global trust while information propagation:
\begin{enumerate}

\item Let be initial credibility of each node as $C$ according to their interest $I$.

\item If a node received a message from a source node, it contacts to its neighbors about the credibility of the source node.

\item If the neighbor nodes know about the credibility of the source node, it renews the credibility score of source node according to information of interest.

\item If the current neighbor does not know the credibility of source node from where information is coming to it then the current neighbor inquires all of its neighbors for their credibility. Each neighbor iterates this procedure, keeping track of the hop count to the source.

\item A threshold value of the credibility will be calculated by considering average credibility score of neighbors of a propagating node, which is used to validate the information at the peer level.
\end{enumerate}

\subsubsection{Decision-making}
A message generating node propagates information according to their credibility and interest of topic. Suppose $C_{info_{t}}$ is propagated at time $t$. If a node has $k$ neighbors, then this information is reached to all $k$ neighbors.
But not all the neighbors have the same interest in the topic of information being propagated. Only those nodes will validate the information who have same interest topic wise. Therefore, a threshold value of the credibility score is set to validate the information and only correct information will be propagated.

When information is propagated then decision on message will be taken based on interest of topic and credibility of nodes,
When a source node propagate information a block of information is created at each time
\begin{equation}
\forall_{info} \in Block(t) : C_{info} = \sum_{k,i}C_{info_{t}}*cred_i
\end{equation}
Where, $C_{info_{t}}$ is a particular type of information generated at time $t$.

To disseminate the trust value for the global trust we used the definition proposed in~\cite{golbeck2009trust} and is given below,

\begin{eqnarray} \label{high-rating}
T_{r,s} = T_{s_{init}}+\frac{\sum_{j \in adj(j)|T_{r,j}} T_{r,j}T_{j,s}}{\sum_{j \in adj(j)|T_{r,j}} T_{r,j}}
\end{eqnarray} 

Where, $T_{r,s}$ is a trust between sender and receiver nodes, $T_{s_{init}}$ is the initial value of trust of a sender node.
The Eq.\ref{high-rating} is used to calculate the trust between two communicating nodes if the information is validated then $T_{s_{init}}$ is updated and trust value of source node is increased. Each receiver node repeats this process, to know the depth of the current node from the source. 
If information is not validated then trust value will be negative and decreased as in Eq.\ref{low-rating}, 

\begin{eqnarray} \label{low-rating}
T_{r,s} = T_{s_{init}}-\frac{\sum_{j \in adj(j)|T_{r,j}} T_{r,j}T_{j,s}}{\sum_{j \in adj(j)|T_{r,j}} T_{r,j}}
\end{eqnarray} 

Finally, validation of information is done by the selected nodes, their credibility for the information is taken as average.
For the set of nodes $V_n$ $\in$ $V$ used for the validation, the credibility score $C_s$ of source node $V_{s}$ according to information of interest $I_{info}$ is the average of the credibility score from nodes in $V_{s}$ by the trust value $T$ from source to each node at peer level:

\begin{equation}
C_{s,{I_{info}}} = \frac{\sum_{i \in V_n} T_{s,i}*C_{I_{info}}}{\sum_{i \in V_n} T_{s,i}}
\end{equation}
Where, $C_{s,{I_{info}}}$ is the credibility score of a node with respective information. If this information is greater than threshold then information will be propagated else blocked. Threshold value will be calculated based on number of nodes interested in particular information.


\subsection{Blockchain based Information Propagation Protocol}

In this section, the information dynamics on social networks is explained. There are three steps in which information propagation will take place. In the first step, a social network e.g. Facebook is used as the underlying network. But in tradition social network any node can send information to its connected nodes, that can be propagated further. To 
make sense with information, that may be right or wrong or 
misinformation(rumor) that should be checked for propagation.

Now a days a third party like Alt news or Cobrapost generally checked the information authenticity on the social network. It takes too much 
time to check the authenticity of information till the information is 
propagated at a large scale.

%

%
Therefore, a system is required which check the authenticity of the information 
implicitly on the network. To do this we have taken the idea from the blockchain 
technology which is used in a financial transaction without using any 
third party,
In our approach, each user in the network is assigned with some 
resources i.e. information. We define two entities that will be 
broadcasted in the network.

\begin{itemize}

\item Node Property(NP) may be defined in form of vector having Node 
ID, Credibility, Info Type.

$NP = \Big([Node\_ID, Credibility, Info\_Type] \Big)$

\item Node service(NS) may be defined as the available message and local 
trust

$NS = \Big(Message, Local\_Trust] \Big)$

\end{itemize}

Blockchain-based information dynamics proposes a new method to propagate information and node identity to peer nodes, allowing each component of the network to verify the information about nodes in the network. 
Blockchain-based information dynamics links cryptographic keys with each $NP$ and $NS$ in the network. We are using the same model of Bitcoin to identify an $NP$ or $NS$ among the network.

\begin{figure}
\begin{center}
\includegraphics[width=.9\linewidth, height=2.0 in]{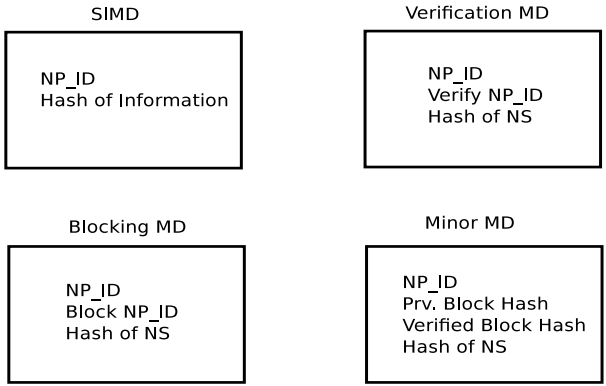}
\caption{Different types of message digest used for verification and 
validation.}
\label{fig7}
\end{center}
\end{figure}

We consider each propagated message as an event like "transaction in 
bitcoin" providing information about the status of a $NP$ and its 
cryptographic information containing $NS$. When a node propagates the 
information, it submits the credibility score and information type in 
network also.
When a node wants to join the network with his $NP$ for the first time, 
it generates a specific information message digest $(SIMD)$ containing 
$NP$ to all nodes. An authentication request is approved when an authenticated $NP$ includes the $(SIMB)$ in a valid block.
Credibility score of the nodes can be updated as per their $NS$ by 
$Verify\_MD$ and $Block\_MD$.
There are 4 types of message digest used in this approach for 
information propagation as shown in Figure \ref{fig7}.

\begin{itemize}
\item SIMD: This message digest is used when a node wants to share information in the network.
\item Minor MD: This message digest is used to validate the blocks to be in the chain.
\item Verification MD: This message digest is used for verification of message.
\item Blocking MD: This message digest tells about the misinformation generated by a node.
\end{itemize}

Note that when blocking information, a $NP / NS$ must provide
a new credibility score to that nodes which has generated the 
information to be in the network.
Miners nodes will try to include $Block\_MD$ and new credibility score 
in the same block to ensure continuity of node status in the network.
If a node wants to propagate information without further verification, 
every node will generate its own message digest
in the network. This will create ambiguity to identify the generator of the message digest. To overcome this issue, every message digest must have a 
hash digest with $NP$ in its entry. This way, the message generator can easily be searched.

\section{Result and Analysis}
In this section, we first explain our experimental settings and next, we discuss the results of our simulations performed using \emph{information propagation using blockchain protocol} considering underlying network topology.

\subsection{Experimental setup}
We have simulated the information propagation using blockchain protocol by using real Facebook network dataset as our underlying networks topology ~\cite{konect:2017:facebook-wosn-wall}:

This network consist of 46,952 nodes and the degree distribution of the network shows the scale-free property. According to interests, Initial credibility is assigned to each person. For example, a person might be more interested in information about movies compared to R\&D. Thus, his credibility will be high for movies information as compared to R\&D.
As per a survey conducted by \textbf{PEW Research Center}, 66\% of social media user discuss the politics, 58\% share and talk about religion, 68\% user share information about science and technology.~\cite{PEW:2018}. Taking inspiration from this \textbf{PEW Research Center} report, the distribution of topics among nodes followed this pattern. 

 Various parameters for simulations are listed in Table\ref{table:1}.

\begin{table}[h!]

\caption{Simulation Parameter}
\centering
\begin{tabular}[5pt]{| c | c |}
\hline

\textbf{Name of Parameter} & \textbf{Value}   \\ [0.8ex] \hline
Nodes & 46952 \\ \hline
$Cred\_Score$ & random (0-1) \\ \hline
$Info\_type$ & [Pol,Tech,Movie,Research]  \\ [1.8ex] \hline
Message generating node & 100 \\ [1.8ex] \hline
\end{tabular}
\label{table:1}
\end{table}

\subsection{Results}
We perform the simulation to explain the proposed secure and trusted information propagation on the social network by using different parameters (see Table 1). We focus on the checking of trusted information on the social network.
We generate the chain of maximum 20 blocks for each message generating nodes. Each block contains the information of nodes and credibility score. Every block is added in chain once it is verified. The result of the simulation is stored in the table as shown in Table \ref{table:2}.

\begin{table}[h!]

\caption{Result}
\centering
\begin{tabular}[5pt]{| c | c | c | c|}
\hline
 \textbf{Noid ID} & \textbf{Info Type} & \textbf{Detection}  \\  \hline
1 & Pol & T \\ \hline
2 & Tech & T \\ \hline
3 & Research & T  \\ \hline
8 & Movie & F \\ \hline
.. & .. & ..  \\ \hline
.. & .. & .. \\ \hline
\end{tabular}
\label{table:2}
\end{table}

\begin{figure}
\begin{center}
\includegraphics[width=1.0\linewidth, height=1.5 in]{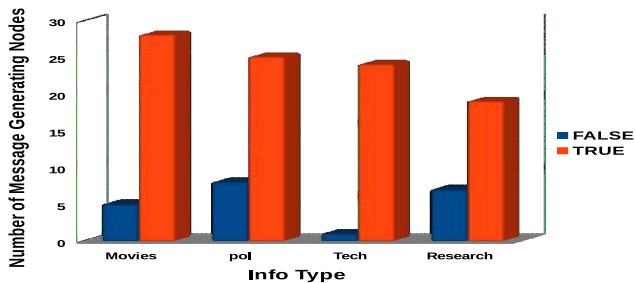}
\caption{{Message generated by nodes}}
\label{verify}
\end{center}
\end{figure}

Finally, we analyze the information authenticity and found that 83\% information is checked correctly that includes different types of messages as shown in figure \ref{verify}. We got the best results for movies, followed by politics, then technology and finally research-related information.  Among all the maximum false are for politics and research. The minimum difference between false and true is observed in the case of research. It also able to identify the message generating nodes. No one can change the identity of the block generated by a particular node due to continuous hashing of blocks in the chain.

\begin{figure}
\begin{center}
\includegraphics[width=1.0\linewidth, height=1.3 in]{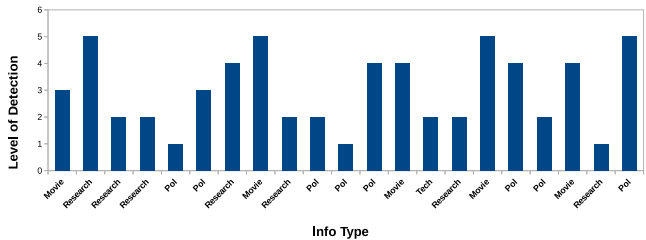}
\caption{{Level of Detection of Information}}
\label{detection}
\end{center}
\end{figure}

We have checked the authenticity of information up to $6^{th}$ hop or level. If the information is validated up to $6^{th}$ hop then information is classified as valid otherwise invalid.  We also found the level of detection for the authenticity of information at the various level as shown in Fig. \ref{detection}. This result shows that the movies related information is checked at the maximum level and the technology related information is checked as early as possible. The average level of detection is also listed in table \ref{table:3}.  The average level of detection for political and research related information is the same, that is detected at level 3. Technology related information is detected at level 2, while the level of detection of movies related information is 4.

\begin{table}[h!]

\caption{Average Level of Detection}
\centering
\begin{tabular}[5pt]{| c | c|}
\hline
\textbf{Info Type} & \textbf{Average Level of Detection}  \\  \hline
 Pol & 3 \\ \hline
 Tech & 2 \\ \hline
 Research & 3  \\ \hline
 Movie & 4 \\ \hline

\end{tabular}
\label{table:3}
\end{table}

%
%
%
%
%
%

\section{Conclusion}
The traditional method of fake information detection is unable to find the source of the message generator in the social network. We have simulated the information propagation using the blockchain protocol. The simulation shows that it is impossible to change the source of the message generator due to the 256-bit hash. Once a message is propagated, it’s authenticity is checked implicitly by using network parameter. Therefore, we no longer require a third party for information verification. We simulated our approach using real Facebook dataset, on which we achieved an accuracy of 83\%.

We plan to include various future directions for this work. This includes the use of additional dynamic networks for our future study. Another direction could be to understand the propagation behaviour of information in the social network by using various larger network datasets.

\section{Acknowledgements}

This work is supported by H2020 framework project, SoBigData, grant number 654024.

\bibliographystyle{IEEEtran}
\bibliography{ref}

\end{document}